\documentclass[aps,prd,reprint,superscriptaddress,showkeys]{revtex4-2}

\usepackage{graphicx}	
\usepackage{amsmath}	
\usepackage{booktabs}   
\usepackage{hyperref}   

\begin{document}

\title{Measuring Peculiar Velocity and Tomographic Redshift Dipole with DESI DR1 Catalogs}

\author{Yi-Wen Wu}
\affiliation{Institute for Frontiers in Astronomy and Astrophysics, Beijing Normal University, Beijing 102206, China}
\affiliation{Department of Physics and Astronomy, Beijing Normal University, Beijing 100875, China}

\author{Jun-Qing Xia}
\email{xiajq@bnu.edu.cn}
\affiliation{Institute for Frontiers in Astronomy and Astrophysics, Beijing Normal University, Beijing 102206, China}
\affiliation{Department of Physics and Astronomy, Beijing Normal University, Beijing 100875, China}

\date{\today}

\begin{abstract}
The so-called ``cosmic dipole tension'' challenges the Cosmological Principle by positing a discrepancy between the Solar System’s peculiar velocity inferred from the Cosmic Microwave Background (CMB) dipole and that derived from large-scale structure number-count dipoles. Here we provide a high-precision determination of the kinematic dipole using the redshift-dipole method applied to the first data release (DR1) of the Dark Energy Spectroscopic Instrument (DESI). By exploiting the Doppler-induced modulation of observed redshifts, this estimator is intrinsically less sensitive to imaging systematics and selection-function uncertainties that can bias traditional number-count measurements. We conduct a tomographic analysis of four tracer populations, Bright Galaxy Sample, Luminous Red Galaxies, Emission Line Galaxies, and quasars, spanning $0.1<z<2.1$. Survey geometry and statistical uncertainties are quantified using 1,000 \texttt{EZmock} realizations. We find that the high-redshift QSO sample implies a peculiar velocity of $v = 357.95_{-48.47}^{+55.05}\,\mathrm{km\,s^{-1}}$, in excellent agreement with the CMB-inferred value of $369.82 \pm 0.11\,\mathrm{km\,s^{-1}}$. By contrast, a complementary number-count analysis yields a significantly enhanced dipole amplitude, which we attribute to leakage of large-scale power and to incompleteness within the DESI DR1 footprint. These results indicate that the redshift dipole provides a cleaner and more reliable probe of the kinematic rest frame, offering strong support for the standard kinematic interpretation at high redshift and helping to resolve the apparent dipole anomaly.
\end{abstract}


\maketitle


\section{Introduction}

The ``cosmic dipole tension'' constitutes one of the most prominent and persistent anomalies confronting the concordance cosmological framework \citep{ABDALLA202249,Kumar_Aluri_2023}. It stems from an apparent inconsistency between the Solar System’s peculiar velocity inferred from the Cosmic Microwave Background (CMB) dipole and that estimated from the dipolar modulation of the large-scale matter distribution. Insofar as statistical homogeneity and isotropy on sufficiently large scales are central to the Cosmological Principle, any genuine mismatch between these two rest-frame determinations would challenge a foundational assumption underpinning the standard $\Lambda$ Cold Dark Matter ($\Lambda$CDM) model. Although the Cosmological Principle is strongly supported by a wide range of observations, recent reports of a dipole amplitude excess, reaching formal significances above $5\sigma$ in some studies \citep[e.g.][]{2022ApJ...937L..31S}, have motivated interpretations invoking either departures from $\Lambda$CDM or, alternatively, residual and as-yet unmodelled observational systematics in large-scale structure (LSS) measurements.

The fiducial reference for our motion is provided by the dipole anisotropy in the CMB temperature field. After subtracting Galactic foregrounds and removing the monopole component ($T_0 \approx 2.726\,\mathrm{K}$), the dominant large-angular-scale feature is a dipole with amplitude $3362.08 \pm 0.99\,\mu\mathrm{K}$ \citep{Planck:2018nkj}. This signal is conventionally interpreted as a purely kinematic dipole produced by Doppler boosting associated with the Solar System barycenter’s peculiar velocity relative to the CMB rest frame. Under this interpretation, the inferred velocity is $v = 369.82 \pm 0.11\,\mathrm{km\,s^{-1}}$ directed towards Galactic coordinates $(l, b) = (264.021^\circ\pm0.011^\circ, 48.253^\circ\pm0.005^\circ)$. This determination plays a central role in precision cosmology and is routinely employed to convert observed redshifts into the cosmological rest frame.

If the Cosmological Principle holds, the LSS should, in an ensemble-averaged sense, define the same kinematic rest frame as the CMB. This expectation enables a stringent internal consistency test, originally articulated by \citet{10.1093/mnras/206.2.377}, who noted that our peculiar velocity should imprint a kinematic dipole in the observed number counts of distant sources. The effect arises from the combined action of relativistic aberration, which redistributes the apparent angular positions of sources and enhances their surface density along the direction of motion, and Doppler boosting, which alters observed fluxes and hence modulates the number of objects crossing a survey’s flux threshold. The resulting number-count dipole is therefore predicted to be aligned with the CMB dipole, with an amplitude that depends on the survey selection (e.g., flux limit) and on the spectral and luminosity-function properties of the tracer population.

For several decades, empirical determinations of the LSS number-count dipole have persistently favoured amplitudes exceeding those expected from the purely kinematic interpretation of the CMB dipole. With the emergence of modern wide-area and effectively all-sky surveys, this mismatch has become more sharply defined and has developed into a statistically significant tension. In particular, studies based on radio-selected sources from the NRAO VLA Sky Survey (NVSS; \citealt{1998AJ....115.1693C}) and the TIFR GMRT Sky Survey (TGSS; \citealt{intema2017gmrt}), as well as infrared-selected quasars from the \textit{CatWISE} catalog \citep{eisenhardt2020catwise}, have repeatedly reported dipole amplitudes a factor of $\sim 2$–$3$ larger than the kinematic expectation \citep{2011ApJ...742L..23S,singal2019large,secrest2021test}. Although the inferred dipole directions are broadly consistent with the CMB dipole, the amplitude excess is difficult to accommodate within standard $\Lambda$CDM and has prompted sustained discussion regarding its origin, ranging from contributions of local structure and bulk flows, to survey-dependent systematics, to a genuine breakdown of statistical isotropy \citep{2022ApJ...937L..31S}.

The physical origin of this persistent discrepancy remains unsettled. Proposed explanations typically fall into two broad classes: (i) complex or residual observational systematics, and (ii) departures from the standard cosmological model \citep{landstrykowski2025cosmicdipoletensionsconfronting}. On the astrophysical side, scenarios invoking large-amplitude “local” contributions posit that unusually large nearby structures, such as a substantial underdensity (void) or overdensity (supercluster), could generate a bulk flow on scales larger than usually assumed \citep{1990ApJ...364..341P,hoffman2017dipole}. More radical possibilities instead attribute the anomaly to new physics, including “tilted” cosmologies in which our Hubble volume samples a superhorizon perturbation \citep{PhysRevD.44.3737,domenech2022galaxy}, or models in which the CMB dipole contains a primordial, intrinsically non-kinematic component \citep{roldan2016interpreting}.

Discriminating among these possibilities, and, crucially, separating them from subtle observational biases, requires an estimator of our peculiar velocity that is fundamentally distinct from number-count–based approaches. Because number-count dipole measurements are potentially vulnerable to low-level systematics associated with survey geometry, calibration, masking, and selection inhomogeneities, an independent and intrinsically cleaner probe is strongly motivated.

The redshift dipole provides such an alternative \citep{singal2025solar,nadolny2021new}. Rather than inferring motion from anisotropies in angular number density, the method exploits the fact that the observer’s peculiar velocity induces a direct Doppler contribution to the observed redshifts of galaxies. As a consequence, the redshift-dipole estimator is largely insensitive to the primary systematics that bias number-density measurements and offers a more direct handle on the kinematic signal. Early applications to Sloan Digital Sky Survey (SDSS; \citealt{ross2020completed,reid2016sdss}) data reported redshift dipoles consistent with the CMB expectation; however, these analyzes were ultimately limited by statistical precision, and thus could not decisively adjudicate the amplitude tension indicated by number-count studies \citep{da2024tomographic,tiwari2024independent}.

In this work, we exploit the unprecedented statistical power of the Dark Energy Spectroscopic Instrument (DESI; \citealt{levi2019dark}) to carry out the most stringent redshift-dipole test of the cosmic dipole to date. We analyze the full set of tracer samples from the first DESI Data Release (DR1; \citealt{abdul2025data}) and combine them to perform a tomographic redshift-dipole measurement across $0.1<z<2.1$. To obtain robust estimates of statistical uncertainties and to assess potential systematic biases, we generate a large ensemble of mock catalogs constructed with the Zel’dovich approximation using the \texttt{EZmock} framework \citep{chuang2015ezmocks}. These realizations enable an accurate treatment of the DESI footprint and survey geometry, as well as the effects of selection functions and other observational systematics. Our goal is to provide a decisive assessment of the reported cosmic dipole anomaly by delivering the tightest constraints from the redshift dipole, thereby testing whether the tension reflects physics beyond the standard model or arises from observational complexities in LSS-based measurements.

This paper is organized as follows. Section~\ref{sec:data} describes the DESI DR1 tracer samples used in this analysis, namely the BGS, LRG, ELG, and QSO catalogs. Section~\ref{sec:method} details the methodology adopted for the redshift- and number-count–dipole measurements and the associated mock-based uncertainty estimation. The resulting dipole constraints from the four tracer samples are presented in Section~\ref{sec:result}. We conclude with a summary and discussion of the implications for the cosmic dipole tension in Section~\ref{sec:discussion}.

\section{Data Release 1 of DESI}\label{sec:data}

The DESI \citep{abdul2025data} is a Stage-IV dark energy experiment mounted on the Nicholas U. Mayall 4-m Telescope at Kitt Peak National Observatory, Arizona. DESI is conducting a five-year spectroscopic survey targeting more than 40 million galaxies and quasars over $\sim 14{,}000\,\mathrm{deg}^2$, yielding an exceptionally large three-dimensional map of the Universe. Its high multiplexing capability is enabled by 5,000 robotic fibre positioners distributed across the focal plane, which allow efficient, wide-field spectroscopy \citep{aghamousa2016desi}.

In this analysis we use observations from DESI DR1 \citep{abdul2025data}, the first major public release comprising the initial year of the main survey. DR1 provides a high-quality dataset for precision cosmology. From DR1 we define four extragalactic tracer samples that together span $0.1<z<2.1$, furnishing the redshift leverage required for a tomographic measurement of the cosmic dipole. The tracer selections follow the criteria described in \citet{2025JCAP...07..017A} and are designed to probe the large-scale structure across distinct cosmic epochs:

\begin{itemize}
    \item \textbf{The Bright Galaxy Sample (BGS):} A low-redshift sample of bright galaxies mapping $0.1<z<0.4$. The catalog contains 300,043 BGS objects. Owing to their high signal-to-noise spectra, BGS redshifts are measured with very high precision.
    \item \textbf{Luminous Red Galaxies (LRGs):} A population of massive, passively evolving galaxies that trace the underlying matter distribution with relatively low bias evolution. We use 2,138,627 LRGs spanning $0.4<z<1.1$ to probe intermediate redshifts.
    \item \textbf{Emission Line Galaxies (ELGs):} Star-forming galaxies selected via prominent [\textsc{o,ii}] emission, providing high number density at higher redshift. Our ELG sample comprises 2,432,072 objects over $0.8<z<1.6$, making it a powerful tracer of structure during this epoch.
    \item \textbf{Quasars (QSOs):} Highly luminous active galactic nuclei that sample the largest cosmological volumes at the highest redshifts accessible in DR1. Although individual QSO redshifts are typically less precise than those of galaxies, their large sample size (856,831) and redshift range ($0.8<z<2.1$) make them indispensable for testing the dipole on the largest scales.
\end{itemize}

\begin{figure*}
	\centering
	\includegraphics[width=2.0\columnwidth]{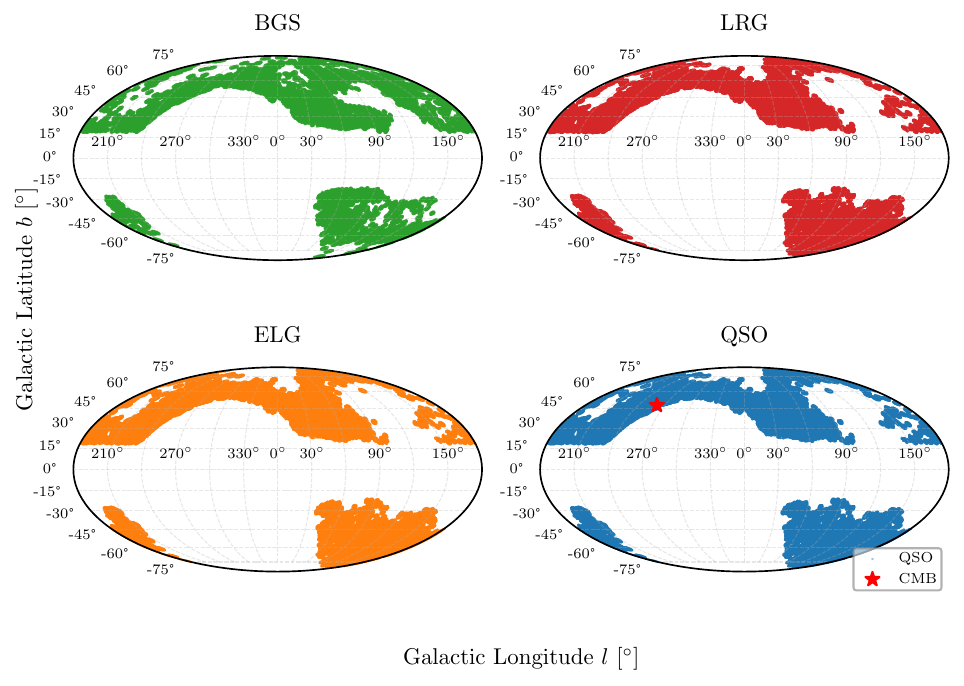}
    \caption{Angular distribution of the four DESI DR1 tracer samples (BGS, LRG, ELG, and QSO) shown in Galactic coordinates $(l,b)$ using a Mollweide projection. Each point corresponds to an individual object in the large-scale structure catalogs. The red star in the lower-right panel (QSO) marks the direction of the CMB dipole, $(l,b)=(264.021^\circ,48.253^\circ)$. The map illustrates the characteristic first-year DESI footprint: several large contiguous survey regions with substantial sky incompleteness. This incomplete angular coverage induces a non-trivial survey window function, giving rise to the window-function effects discussed in Section~\ref{sec:discussion}.}
    \label{fig:footprint}
\end{figure*}

Figure \ref{fig:footprint} shows the angular distribution of the four tracer samples on the sky. The map illustrates the current DESI DR1 footprint, which is separated into the North and South Galactic Caps and is still substantially incomplete relative to full-sky coverage. For reference, we also indicate the CMB dipole direction to facilitate a direct comparison with the LSS dipole measurements presented in this work. The principal characteristics of the tracer catalogs are compiled in Table \ref{tab:info}, including their redshift-estimation uncertainties. These uncertainties incorporate contributions from stochastic measurement scatter, systematic offsets, and a population of catastrophic failures—for example, those induced by emission-line misidentification or contamination and confusion from sky-line residuals \citep{2023ApJ...943...68L}.

\begin{table}
    \caption{Key properties of the DESI DR1 tracer samples used in this analysis. The quoted object counts correspond to the subsets selected for cosmological analyzes in the final DESI large-scale structure catalogs \citep{2025JCAP...07..017A}. Here, $\sigma_v^{\rm std}$ denotes the characteristic redshift measurement precision, expressed as an equivalent line-of-sight velocity uncertainty, while $\sigma_v^{\rm cata}$ is the effective velocity error assigned to the small fraction of objects (0.2--0.7 per cent) affected by catastrophic redshift failures.}
    \label{tab:info}
    \begin{ruledtabular}
    \begin{tabular}{lcccc}
        Tracer & $z$ range & Number & $\sigma_v^{\rm std}$ [$\mathrm{km\,s^{-1}}$] & $\sigma_v^{\rm cata}$ [$\mathrm{km\,s^{-1}}$] \\
        \colrule
        BGS & $0.1-0.4$ & 300,043 & 10 & -- \\
        LRG & $0.4-1.1$ & 2,138,627 & 40--60 & 250 \\
        ELG & $0.8-1.6$ & 2,432,072 & 8--10 & 1000 \\
        QSO & $0.8-2.1$ & 856,831 & 20--125 & 3000 \\
    \end{tabular}
    \end{ruledtabular}
\end{table}

\section{Method}\label{sec:method}

\subsection{The kinematic redshift dipole}

An observer moving with velocity $\vec{v}$ relative to the rest frame defined by the distant LSS will generically measure a dipolar anisotropy in the observed tracer field. In special relativity, radiation emitted at frequency $\nu_0$ by a source observed along direction $\hat{n}$ is Doppler-shifted to an observed frequency $\nu$ according to
\begin{equation}
\nu=\nu_0\frac{1+\hat{n}\cdot\vec{\beta}}{\sqrt{1-\beta^2}}
\equiv \nu_0\delta(\hat{n})~,
\end{equation}
where $\vec{\beta}=\vec{v}/c$ and $c$ is the speed of light. In an expanding universe, the observed redshift receives contributions from both the cosmological Hubble flow and the Doppler effect associated with the observer’s motion. Denoting by $z$ the cosmological redshift, the measured redshift $z'$ satisfies
\begin{align}
1+z'&=(1+z)(1+z_{\rm Doppler}) \nonumber\\
&=(1+z)\frac{\nu_0}{\nu} \nonumber\\
&=(1+z)\delta(\hat{n})^{-1}~,
\label{eq:delta}
\end{align}
where $\hat{n}$ denotes the source direction in the observer’s frame. Strictly, the observed direction differs from the emitted direction due to relativistic aberration. The transformation between the emitted direction $\hat{n}$ and the observed direction $\hat{n}'$ is
\begin{equation}
\hat{n}'=\frac{\hat{n}\cdot\hat{\beta}+\beta}{1+\hat{n}\cdot\vec{\beta}}\hat{\beta}
+\frac{\hat{n}-(\hat{n}\cdot\hat{\beta})\hat{\beta}}{\gamma\left(1+\hat{n}\cdot\vec{\beta}\right)}~,
\end{equation}
with $\gamma=(1-\beta^2)^{-1/2}$ and $\hat{\beta}=\vec{\beta}/\beta$. In the non-relativistic regime relevant for the Solar System, $\beta\ll1$, aberration contributes only at ${\cal O}(\beta^2)$ to the redshift modulation and is therefore negligible for our purposes; we thus adopt $\hat{n}'\simeq \hat{n}$. Expanding Eq.~\eqref{eq:delta} to first order in $\beta$ yields
\begin{align}
1+z'&\simeq (1+z)\left(1-\hat{n}\cdot\vec{\beta}\right) \nonumber\\
z'&\simeq z-(1+z)\hat{n}\cdot\vec{\beta}~,
\label{eq:dipole_approx}
\end{align}
which makes explicit the leading-order dipolar imprint of the observer’s peculiar velocity on the redshift field. The dipole amplitude scales as $(1+z)\beta$, implying an increasing signal-to-noise at higher redshift. Consequently, sources in the direction of motion ($\hat{n}\cdot\vec{\beta}>0$) are observed to be blue-shifted (smaller $z'$), whereas sources in the opposite hemisphere are redshifted relative to the cosmological value.

\subsection{The dipole estimator}

We adopt the tomographic estimator first developed by \cite{da2024tomographic}. Our objective is to constrain the three Cartesian components of the velocity vector $\vec{\beta}$ by fitting the dipolar modulation in Eq.~\eqref{eq:dipole_approx} to the DESI observations. In practice, however, the cosmological redshift $z$ is not a global constant but varies from object to object. We therefore adopt a tomographic strategy, partitioning each tracer sample into a set of narrow spherical redshift shells. 

The choice of the number of redshift bins, $N_{\rm bin}$, reflects a balance between resolving the expected $(1+z)$ scaling of the signal and limiting sensitivity to redshift measurement uncertainties. We set the binning resolution using the typical redshift precision $\sigma_v^{\rm std}$ and the impact of catastrophic redshift failures parameterized by $\sigma_v^{\rm cata}$ (Table~\ref{tab:info}). For the BGS, whose redshifts are exceptionally precise, we employ an effectively continuous tomographic sampling, whereas for tracers with larger statistical scatter and/or higher failure rates (notably ELGs and QSOs) we adopt broader bins to maintain statistical robustness within each shell. Concretely, we use $3000$, $280$, $240$, and $390$ redshift bins for the BGS, LRG, ELG, and QSO samples, respectively.

Within each bin $j$, we approximate the mean cosmological redshift as constant and denote it by $z_j$. The observed redshift of galaxy $i$ in this shell is then modeled as a dipolar perturbation about $z_j$,
\begin{equation}
z'_{ij}(\vec{\beta}) = z_j - (1+z_j)\hat{n}_i\cdot\vec{\beta}~,\qquad i\in\mathcal{S}_j~,
\label{eq:dipole_model}
\end{equation}
where $\mathcal{S}_j$ is the set of galaxies assigned to bin $j$ and $z'$ denotes the measured redshift. 

To estimate $\vec{\beta}$ we construct a weighted estimator that accounts for spatial variations in the DESI selection function. We adopt the total weights provided with the DESI LSS catalogs \citep{2025JCAP...07..017A}, which incorporate completeness corrections, redshift-failure corrections, and mitigation of imaging systematics:
\begin{align}
w'_{\rm tot} &\equiv w_{\rm comp}w_{\rm zfail}w_{\rm imsys}~,\\
w_i &\equiv w_{{\rm tot},i} \equiv {w'_{\rm tot,i}} / {\langle w_{\rm comp}\rangle(n_{\rm tile})}~.
\end{align}

For a trial velocity $\vec{\beta}$, the monopole $z_j$ entering Eq.~\eqref{eq:dipole_model} can be estimated by “de-Dopplering” the observed redshifts. Specifically, inverting Eq.~\eqref{eq:delta} and forming a weighted average over bin $j$ yields
\begin{equation}
z_j(\vec{\beta})=
\frac{\sum_{i\in\mathcal{S}_j} w_i\left[\left(1+z'_i\right)\delta(\hat{n}_i,\vec{\beta})-1\right]}
{\sum_{i\in\mathcal{S}_j} w_i}~,
\label{eq:monopole_est}
\end{equation}
where $\delta(\hat{n}_i,\vec{\beta})$ is the Doppler factor defined above. We then determine $\vec{\beta}$ by minimizing a $\chi^2$-like objective function constructed from the weighted residuals within each redshift shell:
\begin{equation}
\chi^2(\vec{\beta})=
\sum_j
\frac{\sum_{i\in\mathcal{S}_j} w_i^2\left[z'_i-z'_{ij}(\vec{\beta})\right]^2}
{\sum_{i\in\mathcal{S}_j} w_i^2}~,
\label{eq:chi2}
\end{equation}
where the model prediction $z'_{ij}(\vec{\beta})$ is computed via Eq.\eqref{eq:dipole_model} by substituting the monopole $z_j(\vec{\beta})$, which is estimated from the observed redshifts using Eq.\eqref{eq:monopole_est}. The minimizer of Eq.~\eqref{eq:chi2} thus yields the best-fitting velocity vector that consistently describes the global kinematic dipole across the DESI DR1 survey volume.

It is important to note that Eq.~\eqref{eq:chi2} serves strictly as a point estimator to determine the best-fitting parameters. Here, $w_i$ is the total DESI catalog weight defined in Eqs.~(6)--(7), so that the effective normalized weight of object $i$ in shell $j$ is $w_i^2/\sum_{k\in S_j}w_k^2$; these are weights that correct systematics,  rather than inverse-variance weights. The statistical uncertainties on $\vec{\beta}$ are not derived analytically from this least-squares function, but are instead quantified empirically using the variance across the ensemble of 1,000 EZmock realizations (detailed in Section~\ref{sec:EZ}), which naturally accounts for cosmic variance, shot noise, and survey systematics.

\subsection{EZmock and uncertainty estimation}\label{sec:EZ}

Beyond obtaining a best-fitting velocity, reliable uncertainty quantification is essential for any LSS-based inference. In practice, such uncertainties are commonly estimated using synthetic catalogs derived from cosmological simulations, which capture the nonlinear growth of structure and allow controlled injection of survey-specific observational effects (e.g., footprint, selection, and redshift errors) to assess both statistical and systematic contributions. 

The \texttt{EZmock} approach provides an efficient alternative to full $N$-body suites by constructing an approximate density field using the Zel’dovich approximation \citep{zel1970gravitational} and populating it with tracers via a parameterized prescription for galaxy (or quasar) bias. This effective bias model is sufficiently flexible to incorporate linear and nonlinear contributions, as well as deterministic and stochastic components, but must be calibrated against clustering measurements from observations or high-fidelity simulations \citep{chuang2015ezmocks}. In practice, the calibration typically targets summary statistics such as the two-point correlation function (2PCF), the power spectrum, and, when required, higher-order statistics including the bispectrum.

At the inception of this work, the official DESI DR1 mock catalogs were not yet publicly available. Therefore, under the guidance of the primary developers of the DESI mocks, we generated our own suite of mock catalogs for all DESI tracer samples following the exact methodology of \citet{2021MNRAS.503.1149Z}. Consequently, our mock ensemble is statistically equivalent to the official DESI EZmocks, with the specific addition of the injected kinematic dipole required for our pipeline validation.

\begin{table*}
    \caption{Basic parameters of the mock catalogs for the four tracer samples and configuration of the redshift slices adopted for constructing the DESI DR1 \texttt{EZmock} catalogs. The quoted $P_0$ values provide an approximate nominal normalization of the power-spectrum monopole, $P_0(k=0.15\,h\,\mathrm{Mpc}^{-1})$. The mock realizations are generated with number densities and effective volumes larger than those of the corresponding DR1 catalogs; they are subsequently down-sampled to match the observed samples summarized in Table~\ref{tab:info}. For each slice, we generate a cubic realization at the corresponding effective redshift $z_{\rm eff}$ to capture the redshift evolution of both structure growth and tracer bias. Here, $N_{\rm tot}$ denotes the total number of particles in the cubic volume $L^{3}$ prior to imposing the DESI survey geometry and selection.}
    \label{tab:slices}
    \begin{ruledtabular}
    \begin{tabular}{lcccccc}
        Tracer & $n~[h/\mathrm{Mpc}]^3$ & $L~[\mathrm{Mpc}]$ & $N_{\rm tot}$ & $P_0~[\mathrm{Mpc}/h]^3$ & $z_{\rm range}$ (Slice) & $z_{\rm eff}$ \\
        \colrule
        BGS & $3\times10^{-4}$ & 3000 & 8,100,000 & 7000 & $0.1 - 0.4$ & $0.30$ \\
        \colrule
        LRG & $3\times10^{-4}$ & 6000 & 64,800,000 & 10000 & $0.4 - 0.6$ & $0.51$ \\
            & & & & & $0.6 - 0.8$ & $0.71$ \\
            & & & & & $0.8 - 1.1$ & $0.91$ \\
        \colrule
        ELG & $1.5\times10^{-4}$ & 7000 & 51,450,000 & 4000 & $0.8 - 1.1$ & $0.96$ \\
            & & & & & $1.1 - 1.6$ & $1.32$ \\
        \colrule
        QSO & $2.5\times10^{-5}$ & 10000 & 25,000,000 & 6000 & $0.8 - 1.3$ & $1.07$ \\
            & & & & & $1.3 - 1.7$ & $1.50$ \\
            & & & & & $1.7 - 2.1$ & $1.89$ \\
    \end{tabular}
    \end{ruledtabular}
\end{table*}

\subsubsection{Cubic mock catalog generation}

The mock catalogs are constructed assuming a fiducial flat $\Lambda$CDM cosmology with $\Omega_{\rm m}=0.3111$, $\Omega_{\rm b}=0.04897$, $h=0.6766$, $\sigma_8=0.8102$, and $n_{\rm s}=0.9665$, corresponding to the best-fitting parameters from the Planck 2018 analysis \citep{2020A&A...641A...6P}. These parameters enter through the underlying matter power spectrum. In practice, we generate the linear-theory power spectrum $P_{\rm lin}(k)$ using \texttt{CAMB} \citep{lewis2000efficient}. We note that the covariance matrices of two-point clustering observables are typically only weakly sensitive to the precise input spectrum, provided that the mock catalogs reproduce the observed two- and three-point statistics \citep{baumgarten2018robustness}.

The linear matter power spectrum is defined as the Fourier-space variance of the density contrast,
\begin{equation}
P_{\rm lin}(k)\equiv \langle |\delta(\vec{k})|^2\rangle~,
\end{equation}
where $\delta(\vec{k})$ is the Fourier transform of the matter overdensity field. Within Lagrangian perturbation theory, the Zel’dovich approximation gives the linear displacement field $\Psi(\vec{q},t)$ through
\begin{equation}
\nabla_{q}\cdot\Psi(\vec{q},t)=-D_1(t)\,\delta(\vec{q})~,
\end{equation}
with $\vec{q}$ denoting the comoving Lagrangian coordinate and $D_1(t)$ the linear growth factor \citep{zel1970gravitational}. The corresponding Eulerian position of a tracer is
\begin{equation}
\vec{x}(\vec{q},t)=\vec{q}+\Psi(\vec{q},t)~.
\end{equation}
For $\Lambda$CDM, $D_1$ can be evaluated numerically, for example using the integral representation
\begin{equation}
D_1(a)= a^{3}H(a)\frac{5\Omega_{\rm m}}{2}\int_{0}^{a}\frac{{\rm d}{a'}}{{a'}^{3}H({a'})^{3}}~,
\end{equation}
where $a$ is the scale factor and $H(a)$ is the Hubble parameter.

We generate the approximate matter density field with the public implementation of \texttt{EZmock} \citep{chuang2015ezmocks}, which employs the Zel’dovich approximation and a calibrated, parametric prescription to populate tracers. In our setup, four \texttt{EZmock} parameters are tuned to reproduce the observed two- and three-point clustering statistics. The main properties of the mock catalogs are summarized in Table~\ref{tab:slices}, and the calibration procedure is described in Section~\ref{sec:cali}.

\subsubsection{From cubic catalog to light-cone combination}\label{sec:lightcone}

Each \texttt{EZmock} realization is initially generated in a periodic cubic volume at a fixed redshift snapshot, yielding statistically homogeneous clustering within the box. By contrast, the DESI observations exhibit redshift evolution in the clustering statistics, $P(k,z)$, and are further restricted by the survey footprint and veto masks. Constructing realistic DESI-like mocks therefore requires mapping the periodic boxes onto the observed geometry while incorporating the radial selection and (approximately) the light-cone evolution.

To impose the angular footprint, we use the \texttt{MANGLE} package \citep{swanson2008methods} to define the survey mask as a set of spherical polygons, which are then used to trim the mock catalogs to the same sky region as the data. In this step, the observed coordinates $(\mathrm{RA},\mathrm{Dec},z)$ are mapped to Cartesian comoving coordinates $(x_{\rm c},y_{\rm c},z_{\rm c})$ via
\begin{align}
r_{\rm c}(z) &= \int_{0}^{z}\frac{c\mathrm{d}z'}{H_0\sqrt{\Omega_\Lambda+\Omega_{\rm m}(1+z')^{3}}}~,\\
x_{\rm c} &= r_{\rm c}\cos(\mathrm{Dec})\cos(\mathrm{RA})~,\\
y_{\rm c} &= r_{\rm c}\cos(\mathrm{Dec})\sin(\mathrm{RA})~,\\
z_{\rm c} &= r_{\rm c}\sin(\mathrm{Dec})~,
\end{align}
where $r_{\rm c}$ is the radial comoving distance and $H_0$ is the Hubble constant.

To reproduce the observed radial number density, $n(z)$, we down-sample the mock objects as a function of redshift. Specifically, within each redshift bin we randomly remove mock objects with probability $P=1-n_{\rm data}/n_{\rm mock}$. As shown by \citet{2021MNRAS.503.1149Z}, this procedure preserves the clustering statistics of the parent catalog to good approximation. Because the resulting catalog is radially inhomogeneous, we assign each retained mock object an FKP weight to reduce measurement variance,
\begin{equation}
w_{\rm FKP}=\frac{1}{1+n_{\rm mock}P_0}~,
\end{equation}
where $P_0$ is a fiducial value of the power-spectrum monopole. The $P_0$ values adopted for each tracer follow \citet{2025JCAP...07..017A} and are reported in Table \ref{tab:slices}. 

While the above steps yield DESI-like angular and radial selection, a single cubic \texttt{EZmock} box corresponds to a fixed cosmic time and therefore does not encode the redshift evolution of the growth rate $f(z)$, the tracer bias $b(z)$, or the peculiar velocity field across the broad redshift span of the DESI samples. To model this evolution, we implement the light-cone construction described by \citet{2021MNRAS.503.1149Z}. In practice, we use the public code of \texttt{MAKE\_SURVEY} \citep{white2014mock} to execute this procedure. We partition the full redshift range of each tracer into a set of contiguous redshift slices and generate an independent cubic \texttt{EZmock} realization for each slice using the linear power spectrum and growth rate evaluated at an effective redshift, $z_{\rm eff}$. 

For each slice, $z_{\rm eff}$ is computed from the corresponding data sample as the weighted mean redshift,
\begin{equation}
z_{\rm eff}=\frac{\sum_i w_i z_i}{\sum_i w_i}~,
\label{eq:zeff}
\end{equation}
where the sum runs over all objects within the slice and $w_i$ denotes the total catalog weight.

From each cubic realization, we then extract a spherical shell matching the comoving radial bounds of the slice. These shells are subsequently concatenated (“shell stitching”) to form a continuous light-cone volume spanning the full redshift range of the tracer. This procedure ensures that the clustering signal at each redshift is evaluated using the appropriate growth and bias at that epoch. The adopted slice definitions for the DESI DR1 \texttt{EZmock} light-cones are summarized in Table~\ref{tab:slices}. The column $N_{\rm tot}$ denotes the total number of objects generated in the parent cubic volume (prior to any light-cone extraction or survey masking) and is fixed for each tracer type to maintain a consistent baseline number density.

After shell stitching, we apply the DESI angular footprint and veto masks to the full light-cone catalog. Finally, to mimic the specific radial redshift distribution $n(z)$ of the DESI data, we apply the radial down-sampling procedure described in the previous section to the stitched catalog. This ensures that our mock catalogs possess the identical geometry, angular mask, and radial selection function as the DR1 data.

\subsubsection{Random catalog and parameter calibration}\label{sec:cali}

To measure clustering statistics from the mock catalogs in a manner that properly accounts for the survey window function, we generate a matched set of random catalogs. These randoms are processed through the same light-cone construction and survey-geometry pipeline as the mock tracers described in Section~\ref{sec:lightcone}, ensuring full consistency between the estimator, the selection function, and the window function. 

For each tracer and redshift slice, we begin by populating the parent cubic volume with points drawn from a uniform Poisson process. These uniform realizations are then subjected to the identical coordinate transformation, shell extraction and stitching, footprint trimming, and radial down-sampling steps applied to the data-like \texttt{EZmock} catalogs. As a result, the random catalogs reproduce both the angular mask and the radial selection function $n(z)$ of the mocks, and can be used to compute the multipoles of the two-point correlation function, $\xi_{\ell}$, and power spectrum, $P_{\ell}$ (in particular, $\ell=0,2$), required for the calibration procedure.

The \texttt{EZmock} framework maps the underlying matter density field to the tracer distribution through a small set of effective bias parameters (e.g., density-threshold and saturation parameters, and coefficients controlling the PDF mapping). These parameters must be calibrated such that the mocks reproduce the observed clustering amplitude and anisotropy. For each tracer type and redshift slice, we therefore measure the monopole and quadrupole of the correlation function, $\xi_{0,2}(s)$, and the power spectrum, $P_{0,2}(k)$, from the DESI DR1 data, and perform an iterative calibration:
\begin{enumerate}
\item Generate a trial \texttt{EZmock} realization for a candidate parameter set.
\item Measure $\xi_{\ell}$ and $P_{\ell}$ for the realization using the corresponding processed random catalog.
\item Compare the mock statistics to the DR1 measurements.
\item Update the bias parameters and, where applicable, the small-scale velocity-dispersion parameter (to tune Finger-of-God damping in the quadrupole) until the residuals are minimized.
\end{enumerate}
This procedure ensures that the mocks reproduce the DESI two-point clustering signal, capturing both the large-scale bias and the redshift-space distortion (RSD) imprint on intermediate scales.

It is important to note that the raw \texttt{EZmock} realizations are generated in the comoving rest frame of the simulation box and do not inherently contain the observer's peculiar motion. To faithfully reproduce the observational configuration and validate our estimator, we must inject the kinematic dipole signal into the mock light-cones. For each tracer sample, we adopt the best-fitting velocity vector $\vec{\beta}_{\rm data}$ measured from the actual DESI DR1 data (as derived in Section~\ref{sec:result}) as the fiducial input. We modulate the redshifts of all objects in the mock light-cone catalogs according to the Doppler formula (Eq.~\ref{eq:dipole_approx}) using this input velocity. These ``Doppler-shifted'' mocks are then processed through the identical analysis pipeline used for the data. This procedure ensures that our uncertainty estimates account for any potential interplay between the survey window function and the kinematic dipole signal itself.

After determining the optimal parameters, we generate 1,000 independent light-cone realizations for each tracer sample. These mocks form the basis of our statistical error estimation for the dipole measurement. For each realization $k$ ($k=1,\ldots,1000$), we apply the same redshift-dipole pipeline used for the data, minimizing Eq.~\eqref{eq:chi2} to obtain a best-fitting velocity vector $\vec{\beta}_k$. This yields an ensemble ${\vec{\beta}_1,\ldots,\vec{\beta}_{1000}}$ from which uncertainties can be estimated. Because the \texttt{EZmock} initial conditions are drawn from (approximately) Gaussian random fields and the dominant contributions to the measurement uncertainty arise from cosmic variance and shot noise, the recovered velocity components are expected to be well described by a Gaussian distribution. By examining the mock ensemble, we have explicitly verified that the recovered Cartesian velocity components independently follow Gaussian distributions. We therefore adopt, for each component (and for derived quantities such as the dipole direction), the standard deviation across the 1,000 realizations as the corresponding $1\sigma$ uncertainty, and use these values for the results presented in Section~\ref{sec:result}.

\section{Results}\label{sec:result}

In this section, we present our measurements of the Solar System velocity vector $\vec{\beta}$ inferred from the four DESI DR1 tracer samples. The statistical properties of the recovered velocities, quantified using the ensemble of 1,000 \texttt{EZmock} realizations, are summarized in Table~\ref{tab:results}.

\begin{table*}
    \caption{Summary of the kinematic dipole measurements obtained from the DESI DR1 tracer samples. The best-fitting dipole direction, given by the Galactic coordinates $(l,b)$, and the corresponding $68\%$ directional uncertainty $\Delta\theta_{68}$ are inferred from the three-dimensional distribution of velocity vectors across the mock realizations. The dipole amplitude is reported as an equivalent peculiar velocity, with $1\sigma$ uncertainties quoted as the 16th and 84th percentiles of the $\beta$ distribution.}
    \label{tab:results}
    \begin{ruledtabular}
    \begin{tabular}{lccccc}
        Target & Center $l$ [$^\circ$] & Center $b$ [$^\circ$] & $\Delta\theta_{68}$ [$^\circ$] & Median $\beta$ & $v$ [$\mathrm{km\,s^{-1}}$] \\
        \colrule
        BGS & 335.37 & 20.57 & 60.88 & 0.000395 & $118.41_{-47.99}^{+104.10}$ \\
        LRG & 158.21 & 52.79 & 18.54 & 0.000876 & $262.61_{-31.09}^{+34.12}$ \\
        ELG & 337.03 & -73.09 & 20.61 & 0.000734 & $220.05_{-28.39}^{+33.13}$ \\
        QSO & 203.19 & 17.03 & 32.13 & 0.001194 & $357.95_{-48.47}^{+55.05}$ \\
    \end{tabular}
    \end{ruledtabular}
\end{table*}

\subsection{Dipole Amplitude Analysis}

The kinematic dipole amplitude is conveniently parameterized by the dimensionless boost factor $\beta\equiv v/c$. In the mock ensemble, we find that the recovered $\beta$ values are well described by a Gamma-like distribution, as expected for the positive-definite magnitude of a three-dimensional vector whose components are subject to approximately Gaussian fluctuations. In the regime where the signal-to-noise is high and $\beta$ is well separated from zero, this distribution approaches a Gaussian limit.

We observe a clear redshift-dependent evolution of the recovered dipole amplitude across tracer populations. The \textbf{BGS} sample, which probes the local Universe ($0.1<z<0.4$), yields a median velocity of $118.41,\mathrm{km,s^{-1}}$. The comparatively large $1\sigma$ uncertainties ($^{+104.10}_{-47.99}\,\mathrm{km\,s^{-1}}$) indicate that, at low redshift, the kinematic signal is strongly contaminated by sample variance from intrinsic LSS anisotropy and by local bulk flows. 

At intermediate redshifts, the \textbf{LRG} and \textbf{ELG} samples exhibit larger and more stable amplitudes, with $v=262.61^{+34.12}_{-31.09}\,\mathrm{km\,s^{-1}}$ and $v=220.05^{+33.13}_{-28.39}\,\mathrm{km\,s^{-1}}$, respectively. The tightest and most cosmologically informative constraint is obtained from the high-redshift \textbf{QSO} sample ($0.8<z<2.1$), for which we measure $v=357.95^{+55.05}_{-48.47}\,\mathrm{km\,s^{-1}}$. 

This result is in excellent agreement with the CMB-inferred value $v=369.82\pm0.11\,\mathrm{km\,s^{-1}}$, consistent with the expectation that the kinematic contribution dominates the redshift dipole at sufficiently high redshift.
This redshift evolution is physically expected. The measured dipole is a combination of the observer's true kinematic motion and the intrinsic clustering dipole (including local bulk flows). While the kinematic velocity $\beta$ is a global constant, the contamination from large-scale structure (LSS) and bulk flows decays significantly as the survey volume increases. Therefore, “sufficiently high redshift” is quantitatively defined as the regime (typically $z>0.8$, as probed by the QSO sample) where the cosmic variance of the intrinsic LSS dipole and bulk flows drops well below the kinematic signal of$\sim370$ km s$^{-1}$, allowing the kinematic contribution to dominate the measurement.

\subsection{Directional Analysis}

To infer the mean dipole direction $(l,b)$ and quantify its uncertainty, we adopt a three-dimensional vector-averaging procedure. For each mock realization $k$, the recovered direction $(l_k,b_k)$ (with amplitude $\beta_k$) is mapped to a unit Cartesian vector,
\begin{align}
x_k &= \cos b_k \cos l_k, \nonumber\\
y_k &= \cos b_k \sin l_k, \nonumber\\
z_k &= \sin b_k~.
\end{align}
We then compute the component-wise mean over the 1,000 realizations, yielding the mean direction vector $(\bar{X},\bar{Y},\bar{Z})$. The corresponding central direction in Galactic coordinates is obtained by transforming this mean vector back to spherical coordinates,
\begin{align}
l &= \operatorname{arctan2}(\bar{Y},\bar{X}), \nonumber\\
b &= \arcsin(\bar{Z})~.
\end{align}

Standard uncertainty summaries in terms of $(\Delta l,\Delta b)$ can be misleading, particularly at high Galactic latitude where lines of constant longitude converge and a fixed $\Delta l$ corresponds to a latitude-dependent physical separation. We therefore characterize the directional uncertainty using the angular separation $\theta_k$ between each realization and the mean direction. For each realization we compute
\begin{equation}
\cos\theta_k=\sin b\,\sin b_k+\cos b\,\cos b_k\,\cos(l-l_k)~,
\end{equation}
and define the $68\%$ containment radius, $\Delta\theta_{68}$, as the value satisfying $\theta_k\le \Delta\theta_{68}$ for 68\% of the realizations. This statistic is invariant under coordinate singularities and provides a direct, physically meaningful measure of directional precision.

As summarized in Table~\ref{tab:results}, the directional constraints tighten substantially for deeper samples. The BGS measurement yields a broad $\Delta\theta_{68}=60.88^\circ$, consistent with strong contamination from local structure and bulk flows. The LRG sample provides the tightest directional constraint, $\Delta\theta_{68}=18.54^\circ$, with a recovered latitude $b=52.79^\circ$, close to the CMB dipole latitude $b\simeq48.25^\circ$. For the QSO sample, the containment radius is larger ($\Delta\theta_{68}=32.13^\circ$), consistent with its lower effective number density and larger redshift uncertainties, and the recovered direction is $(l,b)=(203.19^\circ,17.03^\circ)$. 

By computing the spherical angular separation between our measurements and the CMB dipole direction ($(l,b)=(264.021^\circ,48.253^\circ)$), we find that the high-redshift QSO sample is consistent with the CMB direction within $1.8\sigma$ (angular separation $\sim58^\circ$ with $\Delta\theta_{68}=32.13^\circ$). The local BGS sample is also consistent within $1.0\sigma$, albeit due to its very large uncertainty. By contrast, the intermediate-redshift LRG and ELG samples show $>3\sigma$ directional deviations. Rather than indicating an unstable rest frame, this redshift-dependent directional scatter is physically expected: at lower and intermediate redshifts, the dipole direction is strongly biased by local bulk flows, intermediate-scale structures, and survey footprint systematics, whereas the true kinematic rest frame is only cleanly recovered at high redshifts (e.g., via QSOs).

\subsection{Statistical Robustness and Mock Verification}

The robustness of our best-fitting parameters, as well as the interpretation of the asymmetric uncertainties, is validated by the extensive suite of mock realizations. As described in Section~\ref{sec:method}, the 1,000 \texttt{EZmock} catalogs generated for each tracer incorporate the DESI DR1 survey geometry, angular and radial selection functions, and the relevant redshift-error model. 

For the QSO sample, the mock-recovered distributions of the dipole amplitude $\beta$ and the dipole direction are close to Gaussian, and their medians closely match the values obtained from the data. This agreement supports the stability of the QSO-based kinematic dipole constraint and confirms that the quoted asymmetric percentiles provide an appropriate summary of the posterior.

By contrast, the LRG and BGS samples exhibit substantially broader directional scatter in $(l,b)$ across the mock ensemble, consistent with the larger uncertainties inferred from the data and reflecting their stronger sensitivity to the incomplete survey footprint and to fluctuations sourced by local and intermediate-scale structure. Overall, the concordance between the data-derived measurements and the corresponding mock distributions indicates that our estimator is effectively unbiased, and that the inferred alignments or residual tensions are unlikely to be artefacts of the analysis pipeline.

\section{Discussion and Conclusion}\label{sec:discussion}

The results in Section~\ref{sec:result} indicate that the redshift dipole inferred from the high-redshift DESI tracers, most notably the QSO sample, is consistent with the kinematic expectation set by the CMB. To place this conclusion in the broader context of the ``cosmic dipole tension'', it is instructive to contrast it with the traditional number-count dipole estimator and to assess the extent to which survey-related systematics can affect density-based measurements. 

Following \citet{xu2022probing}, we therefore carry out a complementary number-count dipole analysis using the same DESI QSO sample. In this framework, the dipole is modeled as a modulation of the observed surface density field, $N(\hat{\boldsymbol{n}})$, induced by the combined effects of Doppler boosting and relativistic aberration. We estimate the dipole parameters by minimizing
\begin{equation}
\chi^2=\sum_i
\frac{\left[N_{\rm obs}(\hat{\boldsymbol{n}}_i)-\bar{N}\left(1+\tilde{A}\cos\tilde{\theta}\right)\right]^2}
{\bar{N}\left(1+\tilde{A}\cos\tilde{\theta}\right)}~,
\end{equation}
where $N_{\rm obs}(\hat{\boldsymbol{n}}_i)$ is the observed number of sources in pixel $i$, $\bar{N}$ is the mean surface density, $\tilde{A}$ is the dipole amplitude, and $\tilde{\theta}$ is the angular separation from the fitted dipole axis. In practice, we pixelize the sky using a \texttt{HEALPix} \citep{gorski2005healpix,zonca2019healpy} map with $N_{\rm side}=64$ (49,152 pixels) and weight each object by the catalog weight $w_{\rm tot}$.

The resulting number-count dipole direction, $(l,b)=(276.64^\circ,44.20^\circ)$, is consistent with the CMB dipole direction. However, the recovered amplitude, $\tilde{A}=0.122$, is far larger than the kinematic expectation, $\tilde{A}\simeq [2+x(1+\alpha)]\beta \sim 10^{-3}$. Anomalously large number-count dipoles have been reported previously in all-sky radio and infrared catalogs (e.g. NVSS, CatWISE; \citealt{secrest2021test}), but the presence of a similarly large excess in a spectroscopic DESI sample motivates a careful examination of potential survey-related contributions.

To diagnose the origin of the excess amplitude, we compute the angular power spectrum $C_\ell$ of the QSO surface-density field using \texttt{NaMaster} \citep{alonso2018unified}, which accounts for mask-induced mode coupling. Following \citet{von2025clustering}, we relate the un-normalized low-$\ell$ amplitudes to the monopole $\mathcal{M}$ and dipole $\mathcal{D}$ via
\begin{align}
C_0 &= 4\pi\,\mathcal{M}^2~,\\
C_1 &= \frac{4\pi}{9}\,\mathcal{M}^2\,\mathcal{D}^2~.
\end{align}
We find a dipole amplitude $\mathcal{D}=0.185$, substantially larger than expected in a statistically isotropic Universe. This strongly suggests that the DR1 QSO density field is affected by significant observational systematics on the largest angular scales. In particular, the small sky fraction $f_{\rm sky}$ and the incomplete characterization of the selection function and completeness at this early survey stage can introduce large-scale power that projects onto the lowest multipoles. In such a regime, the survey window function can induce substantial mode mixing and leakage of power into $\ell=1$, thereby inflating the apparent number-count dipole.

A central outcome of this comparison is the apparent decoupling between the anomalous number-count dipole and the redshift dipole. The number-count estimator is inherently sensitive to spatial variations in the selection function and to large-scale density fluctuations, whereas the redshift dipole in Eq.~\eqref{eq:dipole_model} is driven primarily by the radial Doppler modulation of the redshift values rather than by anisotropies in source density. Consequently, the redshift-dipole estimator is intrinsically more robust to completeness variations and imaging-related systematics that can dominate density-based analyzes. Our results show that even in the presence of strong large-scale power in the angular density field (as indicated by the large $C_1$), the redshift dipole can still recover a kinematic amplitude consistent with the CMB, reinforcing its status as a comparatively clean probe of the kinematic rest frame.

In summary, while the DESI DR1 QSO number-count dipole exhibits an unphysically large amplitude, plausibly driven by limited sky coverage, window-function mode coupling, and residual completeness systematics, the corresponding redshift dipole remains consistent with the CMB rest frame. This contrast supports the interpretation that at least part of the ``dipole tension'' reported in prior count-based studies may originate from unmodeled systematics in source-density maps rather than a failure of statistical isotropy. As DESI continues toward larger area and improved completeness, a joint tomographic analysis combining both redshift- and count-based estimators will provide a substantially more definitive test of isotropy and of our peculiar motion on cosmological scales.

\begin{acknowledgments}
We thank Dr. Cheng Zhao for useful discussions about the usage of \texttt{EZmock}. J.-Q.X. is supported by the National Natural Science Foundation of China, grant No. 12473004, the China Manned Space Program, grant Nos. CMS-CSST-2025-A01 and CMS-CSST-2025-A04; and the Fundamental Research Funds for the Central Universities.
\end{acknowledgments}

\section*{Data Availability}

The data underlying this article are available in the DESI First Data Release (DR1) at \url{https://data.desi.lbl.gov/doc/releases/dr1/}. The \texttt{EZmock} code used to generate the mock catalogs is publicly available at \url{https://github.com/cheng-zhao/EZmock} and \url{https://github.com/cheng-zhao/pyEZmock}. Other software packages used in this analysis, including \texttt{NaMaster} (\url{https://github.com/LSSTDESC/NaMaster}), \texttt{CAMB} (\url{https://camb.info/}), \texttt{HEALPix} (\url{https://healpix.sourceforge.io/}), \texttt{MAKE\_SURVEY} (\url{https://github.com/mockFactory/make\_survey}) and \texttt{MANGLE} (\url{https://space.mit.edu/~molly/mangle/}), are also publicly available.

\bibliographystyle{apsrev4-2}
\bibliography{apssamp}

\end{document}